\documentclass[%
 reprint,
 superscriptaddress,
 amsmath,amssymb,
 aps,
 pre
]{revtex4-1}
\usepackage[latin1 ]{inputenc}
\usepackage[english]{babel}
\usepackage{graphicx}
\usepackage{microtype}
\usepackage{verbatim} %\begin{comment} \end{comment}
\usepackage{bm}
\usepackage{amsfonts}
\usepackage{amsbsy}
\usepackage{color}

\begin{document}

\preprint{APS/123-QED}

\title{Feedback-induced oscillations in one-dimensional colloidal transport}

\author{K.~Lichtner}%, and S.~H.~L.~Klapp} 
\email{lichtner@mailbox.tu-berlin.de}
\affiliation{Institute of Theoretical Physics, Secr. EW~7-1, Technical University Berlin, \\Hardenbergstr. 36, D-10623 Berlin, Germany} 
\author{A.~Pototsky} 
\affiliation{Department of Mathematics, University of Cape Town, Rondebosch 7701, South Africa} 
\author{S.~H.~L.~Klapp}
\affiliation{Institute of Theoretical Physics, Secr. EW~7-1, Technical University Berlin, \\Hardenbergstr. 36, D-10623 Berlin, Germany}

\date{\today}

\begin{abstract}
We investigate a driven, one-dimensional system of colloidal particles in a periodically currogated narrow channel subject to a
time-delayed feedback control. Our goal is to identify conditions under which the control induces oscillatory, time-periodic states.
The investigations are based on the Fokker-Planck equation involving the density distribution of the system. 
First, by using the numerical continuation technique, we determine the linear stability of a stationary density. 
Second, the nonlinear regimes are analyzed by studying numerically the temporal evolution of the first moment of the density distribution. 
In this way we construct a bifurcation diagram revealing the nature of the instability. 
Apart from the case of a system with periodic boundary conditions, we also consider a microchannel of finite length. 
Finally, we study the influence of (repulsive) particle interactions based on Dynamical Density Functional Theory (DDFT).\\
\end{abstract}
\pacs{Valid PACS appear here}
\maketitle

\section{Introduction}
The study of transport of particles in complex geometries is a major topic in nonequilibrium statistical physics with relevance in diverse fields such as biology, condensed matter and nanotechnology
\cite{RevModPhys.81.387,CPHC:CPHC200800526}.
Exemplary systems are colloids in optical (or otherwise modulated) potentials \cite{PhysRevLett.96.190601,PhysRevE.75.060101,C0SM01051K,C2SM07102A}, (bio-)molecules in microchannels \cite{Ros_Eichhorn_Regtmeier_Duong_Reimann_Anselmetti_2005}, cold atoms in optical lattices \cite{PhysRevLett.100.040603}, and magnetic particles adsorbed on ferrimagnetic garnet films \cite{PhysRevLett.105.230602}. Depending on the details of the (often one-dimensional) potential, a variety of fascinating effects can been observed, including ratchet mechanisms \cite{reimann02}, giant diffusion \cite{PhysRevLett.87.010602}, and anomalous (subdiffusive) transport \cite{PhysRevLett.98.020602,PhysRevLett.106.038301,PhysRevLett.103.208301,PhysRevLett.99.060604}. For colloids, which are typically of the size of nano- to micrometer, many of these effects can be monitored by real-space experiments
(see, e.g., \cite{PhysRevE.77.041107,PhysRevLett.93.040603,PhysRevLett.106.168104,NJP103017}). 

In the present paper we investigate a one-dimensional colloidal system where the force exerted by the (static) modulated potential is supplemented by a {\it feedback control} force, i.e., a force depending on the state of the system. Feedback control in the context of Brownian systems is a focus of growing interest, and the method has
already been applied, on a theoretical level, to Brownian
motors \cite{0295-5075-95-5-50003,1742-5468-2011-11-P11016,1742-5468-2011-11-P11016} and flashing or rocking ratchets \cite{0295-5075-81-1-10002}.
Moreover, a first experimental realization of a
feedback-controlled flashing ratchet already exists \cite{PhysRevLett.101.220601}. The overall goal of the feedback control
in this context is to manipulate and/or optimize transport properties such as the current in a flashing ratchet. Beyond this more applicational motivations,
feedback-controlled transport phenomena are also of fundamental interest due to the subtle interplay between state-dependent control protocols, thermodynamics and information theory 
\cite{PhysRevE.85.021104,PhysRevLett.108.030601}.
Very recently, a colloidal system under feedback control was used as experimental realization of an "information heat engine" \cite{Toyabe_Sagawa_Ueda_Muneyuki_Sano_2010}.

Following an earlier study of two of us \cite{EPL40007}, we here employ a feedback control with delay where the control term involves the difference between an
system variable (the control target) at time $t$ and its value
at time $t-\tau$, with $\tau$ being the delay time. Such a time delay often
occurs in experiments due to the time lag between measurement and
feedback. In the general context of control of nonlinear system \cite{schoellhandbook}, time-delayed feedback control (which was introduced by Pyragas \cite{Pyragas1992421})
has been proven to be extremely efficient e.g. for the stabilization
of chaotic orbits; however, steady states can be manipulated as well. In \cite{EPL40007} we used the delayed feedback control 
to manipulate the current of interacting colloids driven through in a tilted washboard potential. Indeed, it turned out that the control can optimize the current and even yield
a current reversal, similar to what has been previously seen for non-interacting systems \cite{PhysRevE.79.041117,PhysRevE.79.041114}.

In the present study we go one step further and ask to which extent the time-delayed feedback control can induce dynamic states {\it not seen} in the uncontrolled
system, which involves a purely static potential. Specifically, we search for the existence of spatio-temporal structures characterized by an {\it oscillatory} distribution of particles. 
That time-delayed feedback control itself can indeed generate
novel dynamics has recently also been seen in other extended systems \cite{PhysRevLett.103.103904}.

Our investigations are based on the numerical solution of the nonlinear Fokker-Planck equation combined with a linear stability analysis. 
Thus, the basic dynamic variable is the time-dependent density field.
Similar to our previous study \cite{EPL40007} our control protocol involves the average particle position as a control target, a quantity which is, in principle, accessible by experiments.
We investigate both, infinite 
systems, that is, colloidal particles on a ring (see \cite{PhysRevE.77.041107} for an experiment), and a finite system to which we refer as microchannel geometry \cite{PhysRevE.80.051103}. 
For both geometries, we do indeed find
oscillatory states at appropriate system parameters and finite values
of the control strength and the delay time.  The character of these oscillations, on the other hand, strongly depends on the set-up.\\
In the last part of the paper, we briefly discuss the impact of repulsive interactions between the particles. This is done 
on the basis of Dynamical Density Functional Theory (DDFT), where the microscopic interactions enter via the free energy functional.
Indeed, in the last years DDFT has been
applied to a variety of driven colloidal systems \cite{0953-8984-21-47-474203}, including attracting colloidal particles in 1D (time-dependent)
ratchet potentials \cite{PhysRevE.83.061401}.

The rest of the paper is organized as follows. Sec.~\ref{Sec.Theory} contains the formulation of the problem. In Sec.~\ref{Sec.DelayInstability}, we determine the stability of a nontrivial stationary, spatially-periodic density distribution by linearizing the Fokker-Planck equation subject to a time-delayed control force. For the nonlinear regimes, we provide numerical solutions of the (full) Fokker-Planck equation and construct a bifurcation diagram based on monitoring the amplitude of the solutions. In Sec.~\ref{SecDYNMicro}, we discuss the impact of the channel geometry on the oscillation instability. For that purpose, we consider a microchannel of finite length $L$ and follow the same procedure as for the infinite system. 
The role of interactions between the particles is briefly discussed in Sec.~\ref{SecPartInteract}.

\section{Theory}\label{Sec.Theory}
Our model system consists of overdamped colloidal particles in a one-dimensional channel of length $L$. The particles are subject to a spatially periodic, symmetric ``washboard'' potential $U_\text{wb}(z)=U_0\cos^2(kz)$, where $k$ defines the wavelength and $U_0$ is the amplitude. Henceforth we set $k\sigma=1$ where $\sigma$ corresponds to the effective (gyration) radius of the particle. The washboard potential is tilted by a constant force $\mathbf{F}_{\text{bias}}=F_0\hat{\mathbf{z}}$ (with $\hat{\mathbf{z}}$ being the unit vector in z-direction) 
corresponding to a linear potential $U_\text{bias}=-F_0z$. The tilting leads to an effective motion of the particles in the direction of sign$(F_0)$ along the $z$-axis. 
In addition to these static potentials, we assume that the particles are subject to a time-delayed feedback control force of the form \cite{EPL40007}
\begin{align}
F_\text{fb}(t,\tau)=-K_0(1-\tanh\left[\bar{f}(t)-\bar{f}(t-\tau)\right]).\label{eq.fbforce}
\end{align}
where $\bar{f}(t)$ is a space-averaged coupling function which depends on the internal dynamical variables of the system. Specifically, we set 
\begin{align}
 \bar{f}(t)=\int_{z_0}^{L+z_0}f(z)\rho(z,t)\,dz\label{eq.fbcoupling},
\end{align}
with $z_0$ being the coordinate of the origin of the z-axis.
Note that when the feedback term is switched on, the effective constant driving force is given by $\gamma=F_0-K_0$.

In Ref.~\onlinecite{EPL40007}, we used a force of the type (\ref{eq.fbforce}) to manipulate, or, more precisely, to revert the net current induced by the biasing force. Here, we aim to explore to which extent such a feedback force can
induce oscillatory (time-periodic) density states, which correspond to synchronized oscillations of the particles along the channel. 
We note that the control force in Eq.~(\ref{eq.fbforce}) is of Pyragas type \cite{Pyragas1992421}, i.e. the difference of the ``control target" at time $t$ and its value at time $t-\tau$ is used as input for the feedback loop, where $\tau$ is called ``delay time" and $K_0$ is the control amplitude. 
We stress that the ansatz for the feedback force in Eq.~(\ref{eq.fbforce}) is clearly of heuristic nature, and thus cannot be derived from any physical potential.

Collecting all external contributions gives the total external potential
\begin{align}
 U_{\text{ext}}=&U_\text{wb}(z)+U_\text{fb}(t,\tau)+U_\text{bias}(z)\nonumber\\
=&U_0\cos^2(z)+K_0 z(1-\tanh\left[\bar{f}(t)-\bar{f}(t-\tau)\right])-F_0z,\label{eq.extpot}
\end{align}
where we assumed that the contribution of the feedback force is linear in the position coordinate $z$.
In the absence of interactions between the particles, their Brownian motion is governed by the following Fokker-Planck equation \cite{risken}
\begin{align}
\Gamma^{-1}\frac{\partial \rho(z,t)}{\partial t}=&k_BT\dfrac{\partial^2 \rho(z,t)}{\partial z^2}-\dfrac{\partial }{\partial z}\left[\rho(z,t)\tilde{\mu}(z,t;\tau)\right]\label{eq:FPl1}, 
\end{align}
where the drift coefficient $\tilde{\mu}$ can be calculated from the external potential $U_\text{ext}$ via $\tilde{\mu}=-U_\text{ext}'$ where $'$ denotes the derivative w.r.t. $z$. The mobility coefficient in Eq.~(\ref{eq:FPl1}) is related to the diffusion constant via $\Gamma=\beta D_0$ [where $\beta=1/(k_BT)$] and we set its value to $\Gamma \tau_\mathrm{B}k_\mathrm{B}T/\sigma^2=1$. 
Time is measured in units of the Brownian time scale $\tau_\mathrm{B}=\sigma^2/(\Gamma k_BT)$, which is of the order of $10^{-9}$s for typical Brownian particles.

In the following, we choose a positive value for $F_0$ such that the particles move preferentially to the right when the control force is being switched off. Also, for the main part of our investigations, we consider the system to be infinitely extended (``circular ring''),
that is, we apply periodic boundary conditions over a length $L$, which is a multiple of a period of the washboard potential, i.e. $L=n\pi\sigma$, with $n=1,2,\dots$. However, in Sec.~\ref{SecDYNMicro}, we also discuss the impact of finite channel length.

\section{Delay-induced instability}\label{Sec.DelayInstability}
In the absence of control ($K_0=0$), the driven system ($F_0>0$) settles into a stationary, non-oscillatory state. The corresponding distribution $\rho_s(z)$
can be found analytically \cite{reimann02}. In the presence of control, the behavior of the system depends on the interplay
of the parameters of the control term, on the one hand, and the tilted washboard potential, on the other hand.
For certain parameter combinations we still find a stationary state, in which the difference $\bar{f}(t)-\bar{f}(t-\tau)$ disappears.
However, as we will demonstrate, for suitable parameters the stationary distribution may become unstable, leading to a stable
time-periodic distribution $\rho(z,t)$. To interpret this instability, it is useful to reconsider
Eqs.~(\ref{eq.fbforce}-\ref{eq.fbcoupling}) from a somewhat different perspective. In particular, since the control target
$\bar{f}(t)$ involves a spatial integral over the entire distribution $\rho(z,t)$, we have effectively
introduced a coupling between the colloidal particles. Moreover, this coupling is of infinite range (within the periodic system considered).
Thus, the time-delayed feedback control introduces an effective interaction of mean-field type, and it is this interaction, which may
lead to new stationary states as well as new dynamic regimes, associated with time-periodic density oscillations. We note that the particular type of feedback used here is of purely heuristic nature. This implies that the
Fokker-Planck equation Eq.~(\ref{eq:FPl1}) cannot be directly translated into a corresponding system of coupled Langevin equations.

\subsection{Linear stability analysis}\label{SecLSA}
We start by applying a linear stability analysis in order to investigate the impact of the delayed feedback control. To this end, we rewrite 
the dimensionless Fokker-Planck equation in terms of the effective potential $U_{\rm eff} = U_0 \cos^2{z} + (K_0 - F_0)z$ as follows
\begin{align}
\frac{\partial \rho(z,t)}{\partial t}=&\dfrac{\partial^2 \rho(z,t)}{\partial z^2}+\dfrac{\partial }{\partial z}\left[\rho(z,t)\frac{\partial U_{\rm eff}}{\partial z}\right]\nonumber\\
&-K_0\tanh\left[\bar{f}(t)-\bar{f}(t-\tau)\right]\dfrac{\partial\rho(z,t)}{\partial z}\label{eq:FPls}.
\end{align}
 The non-trivial stationary state $\rho_s(z)$ satisfies
\begin{align}
\label{stat}
0=\frac{\partial}{\partial z}\left(\rho_s\frac{\partial U_{\rm eff}}{\partial z}+\dfrac{\partial \rho_s(z)}{\partial z}\right),
\end{align}
which can be written as an eigenvalue problem
\begin{align}
\label{eig_stat}
\hat{L}\rho_s = \lambda \rho_s,
\end{align}
with the stationary Fokker-Planck operator $\hat{L}=\partial^2_z+(\partial_z U_{\rm eff})\partial_z+\partial_z^2U_{\rm eff}$ and zero eigenvalue $\lambda=0$.

We are interested in the onset of an oscillatory instability of $\rho_s$. Following the approach, developed earlier \cite{Pot09,P12}, we set
\begin{align}
\label{ansatz}
\rho(z,t)=\rho_s(z) + \epsilon\left(C(z)\cos{\omega t}+S(z)\sin{\omega t}\right),
\end{align}
where $\epsilon$ is the (small) amplitude of the perturbation, $\omega$ is the unknown onset frequency and
the unknown functions $C(z)$ and $S(z)$ determine the shape of the perturbation. With this
ansatz, the Eq.\,(\ref{eq:FPls}) can be linearized in $\epsilon$ to yield
\begin{align}
\label{lin_eq1}
-\omega C \sin{\omega t}+\omega S \cos{\omega t}=&C^{\prime\prime}\cos{\omega t}+S^{\prime\prime}\sin{\omega t} \nonumber\\
+&\partial_z\left[U_{\rm eff}^\prime(C\cos{\omega t}+S\sin{\omega t})\right]\nonumber\\
-&K_0\rho_s^\prime\left[\bar{f}(t)-\bar{f}(t-\tau)\right],
\end{align}
where $\prime$ stands for the derivative w.r.t. $z$ and the perturbed mean field is given by
\begin{align}
\label{mf}
\bar{f}(t)=\langle C\rangle\cos{\omega t}+\langle S\rangle\sin{\omega t},
\end{align}
with $\langle C\rangle =\int_{z_0}^{L+z_0}f(z)C(z)\,dz$ and $\langle S\rangle =\int_{z_0}^{L+z_0}f(z)S(z)\,dz$.
The initial moment of time can always be chosen in such a way that, for instance, $\bar{f}(t) =\cos{\omega t}$. This implies two additional integral conditions on the functions $C(z)$ and $S(z)$
\begin{align}
\label{int_cond}
\int_{z_0}^{L+z_0}f(z)C(z)\,dz=1,\,\,\,\int_{z_0}^{L+z_0}f(z)S(z)\,dz=0.
\end{align}
Then the difference $\bar{f}(t)-\bar{f}(t-\tau)$ becomes
\begin{align}
\label{mf_diff}
\bar{f}(t)-\bar{f}(t-\tau) = \cos{\omega t}(1-\cos{\omega \tau})-\sin{\omega t}\sin{\omega \tau}.
\end{align}
Finally, equating the coefficients of $\sin{\omega t}$ and $\cos{\omega t}$ in Eq.\,(\ref{lin_eq1}), we obtain two coupled
equations for the unknown functions $C(z)$ and $S(z)$
\begin{align}
\label{lin_eq2}
S^{\prime\prime}=&-\omega C-\partial_z\left[U_{\rm eff}S\right]-K_0\sin{\omega\tau}\rho_s^\prime\nonumber\\
C^{\prime\prime}=&\omega S-\partial_z\left[U_{\rm eff}C\right]+K_0(1-\cos{\omega\tau})\rho_s^\prime.
\end{align}
In order to find the stability threshold, one needs to solve Eqs.\,(\ref{lin_eq2}) simultaneously with
Eq.\,(\ref{eig_stat}). To this end we proceed as follows.
We rewrite Eqs.\,(\ref{eig_stat},\ref{lin_eq2}) as an autonomous dynamical system of seven first order equations,
including the equation for $z$, which takes the form $z^\prime=1$. The total number of the system parameters is thereby extended by two
additional parameters, namely by the onset frequency $\omega$ and the (zero) eigenvalue $\lambda$. The
above dynamical system of seven equations is supplemented with three integral conditions
on the functions $\rho_s(z)$, $C(z)$ and $S(z)$. These are given by Eqs.\,(\ref{int_cond}) and by the normalization
condition on $\rho_s$
\begin{align}
\label{norm}
\int_{z_0}^{L+z_0}\rho_s(z)\,dz=N,
\end{align}
where $N$ is the normalization parameter.
\begin{figure}[htpb!]
    \begin{center}
    \resizebox{85mm}{!}{\includegraphics{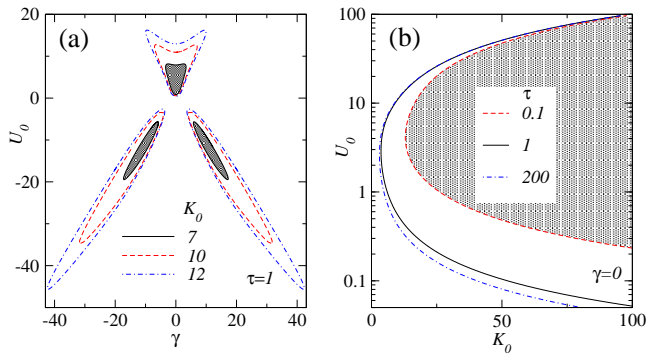}}
      \caption{(Color online) (a) Stability threshold in the plane $(\gamma,U_0)$, obtained for $\tau=1$ and different values of $K_0$, as given in the legend. The stationary state is unstable in the area enclosed by the respective curves (shaded area for $K_0=7$). (b) Stability threshold in the plane $(K_0,U_0)$ for zero drive $\gamma=0$ and different delay times as in the legend. The instability region always lies to the right from the respective curve (shaded area for $\tau=0.1$)}
      \label{F2}
    \end{center}
\end{figure}
The boundary conditions for all
involved functions are taken to be periodic in the interval $z\in[z_0,z_0+L]$. This boundary value problem (BVP) of seven
equations and three integral conditions is then solved using the numerical continuation
technique (AUTO) \cite{AUTO} (see Appendix\,\ref{append} for details).

\subsection{Stability thresholds}

Before proceeding, it is important to notice that the linear stability of the stationary distribution $\rho_s$ crucially depends on the choice of the coupling function $f(z)$ [see Eq.(\ref{eq.fbcoupling})]. Thus, the solution of the BVP Eqs.\,(\ref{lin_eq2}) is invariant under the shift of the coordinate system $z\rightarrow z + \delta$, with arbitrary $\delta$, only if the coupling function $f(z)$ is itself periodic with the period $L$. Indeed, the integrals $\int_{z_0}^{z_0+L}f(z)C(z)\,dz$ and $\int_{z_0}^{z_0+L}f(z)S(z)\,dz$, are shift-invariant if $f(z)$ is $L$-periodic.  On the contrary, if the period of $f(z)$  is different from $L$, or if $f(z)$ is a non-periodic function, then the stability threshold depends on the particular choice of origin of the $z$-axis, i.e. it depends on $z_0$.
Following \cite{EPL40007} we use a linear, non-periodic coupling function $f(z)=z$. For further calculations, the origin of the $z$-axis is chosen in the maximum of the washboard potential $U(z)$, implying that $z_0=0$.

First, we fix $\tau=1$ and compute the stability threshold in the plane of parameters $(\gamma,U_0)$, where $\gamma=F_0-K_0=0$, for three different values of the coupling strength $K_0=7,\,10,\,12$, as shown in Fig.\,\ref{F2}(a). The stationary density, normalized with $N=1$, i.e. $\int_0^{\pi} \rho_s(z)\,dz=N=1$, is unstable in the regions bounded by the corresponding closed curves. Thus, for $K_0=7$, the instability occurs in the shaded area. As expected, the area of the instability expands if the coupling strength $K_0$ is increased. 

Interestingly, the stability diagram for the case when the origin of the $z$-axis is chosen in the minimum of the washboard potential ($z_0=\pi/2$), can be obtained from Fig.\,\ref{F2}(a) by the transformation $U_0\rightarrow -U_0$. For any other choice of $z_0$, the topology of the stability threshold is much more complex and generally contains four different bounded regions (not shown).
\begin{figure}[htpb!]
    \begin{center}
   \resizebox{74mm}{!}{\includegraphics{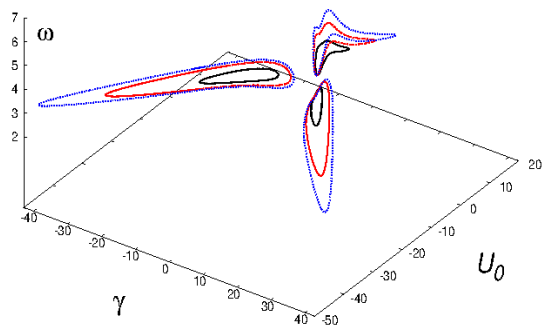}}
        \caption{(Color online) Three-dimensional view of Fig.\,\ref{F2}(a). The onset frequency $\omega$ as a function of $U_0$ and $\gamma$  for $\tau=1$ and three different $K_0$. The projection onto the $(\gamma,U_0)$ plane recovers Fig.\,\ref{F2}(a). }
      \label{F3}
    \end{center}
\end{figure}

The effect of the time delay is demonstrated in Fig.\,\ref{F2}(b) for the choice $\gamma=0$. The stationary density is unstable in the area, which always stretches towards larger values of the coupling strength $K_0$, as shown by the shaded area for $\tau=0.1$. Decreasing $\tau$ leads to the suppression of the instability, which clearly demonstrates that the instability is induced by the presence of the time delay in the coupling term.
 
However, it should be emphasized that having only a time-delayed coupling does not suffies to induce the instability. Thus, from Fig.\,\ref{F2}(b) it follows that even if $\tau$ and $K_0$ are rather large, e.g. $\tau=200$ and $K_0 \approx 100$, the stationary density is linearly stable for a vanishingly weak or an infinitely strong washboard potential $U_0$. At fixed $K_0$, only a certain combination of $\tau$ and $U_0$ renders the system unstable with respect to an oscillatory perturbation. Consequently, the onset of the synchronized time-periodic state is a the effect of the combined action of the time-delayed coupling and stationary periodic external modulation in the form of the washboard potential.

By following the stability threshold in the parameter space, we additionally obtain the onset frequency $\omega$ directly on the threshold. The latter carries an important information about the time scale of the newly born oscillatory states, given by $T=2\pi/\omega$. The three-dimensional view of Fig.~\ref{F2}(a), extended by the onset frequency $\omega$, is shown in Fig.~\ref{F3}. It can be seen that the smallest temporal period $T\sim 1$ corresponds to positive $U_0$, whereas the period of the time-periodic states born at negative $U_0$ is up to three fold larger, i.e. $T\sim 6$.

\subsection{Nonlinear regime: Numerical study of the one-body distribution}\label{SecDYNNA}
\begin{figure*}[tbp!]
    \begin{center}
      \resizebox{140mm}{!}{\includegraphics{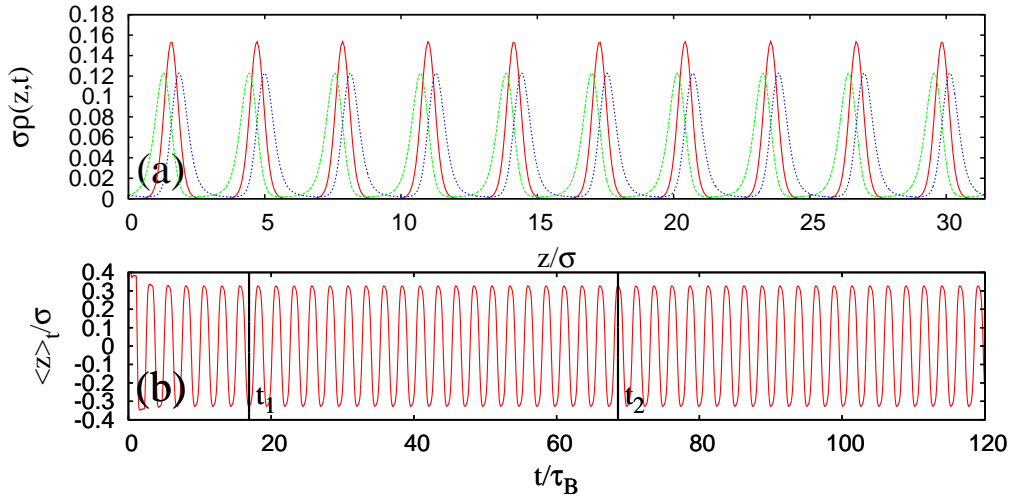}} %180
      %\resizebox{140mm}{!}{\includegraphics{U7K7F7prb.eps}} %180
      \caption{(Color online) Results for a controlled system in the periodic regime. (a) Density distribution $\rho(z,t)$ as a function of position for selected times $t_0/\tau_B=0$ (red curve), $t_1/\tau_B=16.9$ (green-dashed curve) and $t_2/\tau_B=68.6$ (blue-dashed curve). (b) Average particle position $\langle z\rangle_t$ as a function of time. The parameters are $U_0=7k_BT$, $F_0=7k_BT/\sigma$, $K_0=7k_BT/\sigma$, $\tau=\tau_B$ and $L=10\pi\sigma$.}
      \label{fig.CircleDensityUnstable}
    \end{center}
\end{figure*}
The observation of the linear instability with respect to oscillatory perturbations predicts possible deviations from the stationary state for a given (``overcritical") parameter set.
In order to study the full nonlinear dynamics, however, the full Fokker-Planck equation of the system has to be solved. To explore these nonlinear effect we solve Eq.~(\ref{eq:FPl1}) numerically. Specifically, we employ a standard ``Forward-Time Centered-Space" (FTCS) finite difference method \cite{Press:2007:NRE:1403886} and integrate Eq.~(\ref{eq:FPl1}) starting from an inital distribution $\rho(z,t=0)$.

As initial density distribution we choose the equilibrium density distribution in a one-dimensional washboard potential with periodic boundary conditions and the external bias, as well as the control, being switched off. This implies
\begin{align}
\rho(z,t=0)=\rho_0\exp\left[-\beta U_0\cos^2\left(\dfrac{z}{\sigma}\right)\right],\label{Eq.EquiPRB}
\end{align}
where $\rho_0$ ensures the normalization condition $\int_{0}^{L} \rho(z,t=0) dz=N=1$.
We consider a fixed control amplitude $K_0=7k_\mathrm{B}T/\sigma$ and focus on parameter values near the stability threshold [see Fig.~1(a)]. Specifically, we consider 
the ``balanced case" $\gamma=F_0-K_0=0$ and washboard amplitude values of $U_0=8k_BT$ (linearly stable) and $U_0=7k_BT$ (linearly unstable), respectively. Beginning with the latter case, 
we plot in Fig.~\ref{fig.CircleDensityUnstable}(a) snapshots of the density distribution $\rho(z,t)$ for three subsequent times. The initial distribution $\rho(z,t=0)$ is periodic in the position coordinate $z$ with a (spatial) period that is equal to the valley-to-valley distance of the washboard potential, that is $\lambda_\text{wb}=\pi\sigma$.
As expected, the values for the density distribution are increased at the position coordinates $z_i^\text{valley}=i\pi\sigma$ (where $i=[0,1,2,3,\dots])$ corresponding to the valley positions of the washboard potential.
The specific times $t_1$, $t_2$ are chosen such that the appertaining density distributions (shown as the green-dashed curve and the blue-dashed curve in Fig.~\ref{fig.CircleDensityUnstable}(a), respectively) have a maximum displacement from the initital configuration.
Inspecting the curves, we find that after a response time of roughly $5\tau_B$, the system settles indeed into a stable time-periodic density state, i.e. $\rho(z,t+T)=\rho(z,t)$, where the distribution oscillates around a washboard minimum position with a maximum displacement of approximately $0.33\sigma$.
The appearance of such stable oscillations is consistent with the stability diagram in Fig.~\ref{F2}(a). 
We supplement the discussion by plotting in Fig.~\ref{fig.CircleDensityUnstable}(b) the particle position averaged over one period of the potential, that is, $\langle z\rangle_t=\int_0^{\lambda_\text{wb}}\rho(z,t)dz$.
\begin{figure*}[tbp!]
    \begin{center}
    \resizebox{140mm}{!}{\includegraphics{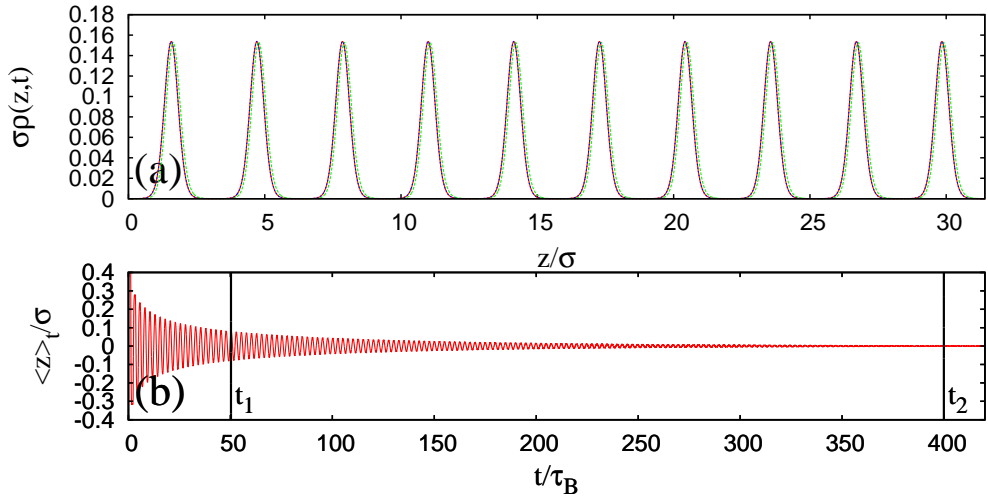}} %180
      %\resizebox{140mm}{!}{\includegraphics{U8K7F7prb.eps}} %180
      \caption{(Color online) Results for a controlled system in the regime where the final state is stationary. (a) Density distribution $\rho(z,t)$ as a function of position for selected times $t_0/\tau_B=0$ (red curve), $t_1/\tau_B=50.4$ (green-dashed curve) and $t_2/\tau_B=400.0$ (blue-dashed curve). (b) Average particle position $\langle z\rangle_t$ as a function of time. The parameters are $U_0=8k_BT$, $F_0=7k_BT/\sigma$, $K_0=7k_BT/\sigma$, $\tau=\tau_B$ and $L=10\pi\sigma$.}
      \label{fig.CircleDensityStable}
    \end{center}
\end{figure*}
Clearly, $\langle z\rangle_t$ oscillates as a function of time with the same frequency $\omega$ as the frequency of the density oscillations. From now on, we therefore use the function $\langle z\rangle_t$ to obtain the cycle time $T=2\pi/\omega$.

We now turn to the case $U_0=8k_BT$. For this parameter, the perturbating forces, specifically the constant tilting force $\mathbf{F}^\text{bias}$ and the feedback force $\mathbf{F}^\text{f.b.}$, do not lead to an oscillating state. This is illustrated in Fig.~\ref{fig.CircleDensityStable}. It is seen that the oscillations at early times 
are being damped resulting in a stationary, non-periodic density for times $t\gtrsim 400\tau_B$. 
In Fig.~\ref{fig.CircleDensityStable}(a) we show the results for the stationary density profile as a blue-dashed curve, as well as for two additional profiles, where the red and green-dashed curves represent the initial and one intermediate state, respectively. As can be seen from Fig.~\ref{fig.CircleDensityStable}(b), the average particle position $\langle z\rangle_t$, converges to a constant value $\langle z\rangle_\text{stat}/\sigma\approx 0$. In other words, the density displacement related to the equilibrium position at each potential valley vanishes and, thus, the profiles for times $t_2/\tau_B=400$ and $t_0/\tau_B=0$ coincide.

We note that for both values of $U_0$ considered (cf. Figs.~\ref{fig.CircleDensityUnstable}-\ref{fig.CircleDensityStable}), the dynamics at early times is transient. In the oscillatory case, the transient regime lasts for about $t_\text{res}\approx 5\tau_B$. For the linear stable case, the transient response time can be much larger. For example, for the parameter values that we considered in Fig.~\ref{fig.CircleDensityStable} the stationary state is not reached for times less than $t_\text{res}\approx 400\tau_B$.

\begin{figure}[tbp!]
    \begin{center}
     \resizebox{70mm}{!}{\includegraphics{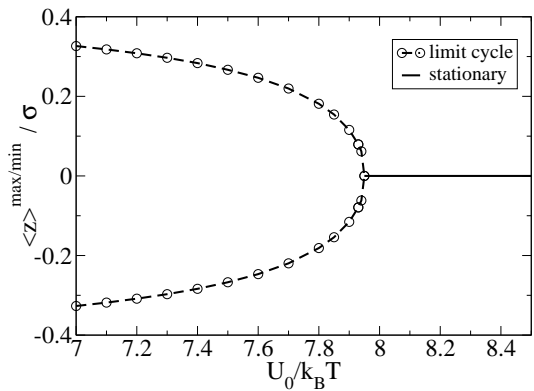}}
      %\resizebox{120mm}{!}{\includegraphics{bif_circle.eps}}
      \caption{Local extrema values of the function $\langle z\rangle_t$ as a function of the washboard amplitude $U_0$. The parameters are $F_0=7k_BT/\sigma$, $K_0=7k_BT/\sigma$, $\tau=\tau_B$, $L=10\pi\sigma$.}
      \label{Fig.circlebif}
    \end{center}
\end{figure}

We have repeated the numerical calculations described above for a range of parameters $U_0$ and the choice $\gamma=F_0-K_0=0$. In this way we can
construct a bifurcation diagram characterizing the nature of the instability. As a measure of the instability, we use the oscillation 
amplitude of $\langle z\rangle_t$. Specifically, we obtain local extrema values $\langle z\rangle^\text{max/min}$ from the function $\langle z\rangle_t$ and average these over several periods.
The results are summarized in Fig~\ref{Fig.circlebif}.
At large values of $U_0$ we only find one stable attracting fixed point $\langle z\rangle^\text{max/min}=0$.
However, by decreasing the washboard amplitude $U_0$ the stationary state (characterized by $\langle z\rangle^\text{max/min}=0$ for all times $t$ greater than the transient time) loses stability and stable limit cycle oscillations 
occur (see, e.g., Fig.~\ref{fig.CircleDensityUnstable}). This happens in an essentially continuous manner, as Fig.~\ref{Fig.circlebif} reveals. We thus conclude that, upon decreasing $U_0$ below $U_0^c\approx 7.95k_BT$ the system undergoes a supercritical Hopf bifurcation. For $U_0<U_0^c$ all resulting trajectories perform limit cycle oscillations about the former stationary state $\langle z\rangle^\text{max/min}=0$. 
We note that all neighboring trajectories approach the limit cycle. Thus, for $U_0<U_0^c$, the limit cycle is stable and the only attractor in the system.
Furthermore, due to the spatially left-right symmetry in the potential, the local extrema of $\langle z\rangle_t$ appear in symmetrical pairs at $\pm|\langle z\rangle^\text{max}|$.

\subsection{Cycle time for the oscillating density state}\label{SecPerCycle}
An interesting question is to which extent the density oscillation frequency depends on the different system parameters such as washboard potential amplitude $U_0$ and time delay $\tau$. 
In Fig.~\ref{F3} we have already shown results for the
{\it onset} frequency based on the linear stability analysis. Here we present
corresponding data obtained in the nonlinear regime. To this end we define the cycle time $T$ as the overall travel time
 for a full maximum displacement of the density distribution (towards and back). As argued before, this time can be obtained from the average particle position $\langle z\rangle_t$ by measuring the (time) distance between two maxima [see Fig.~\ref{fig.CircleDensityUnstable}(b) for an example]. 
 \begin{figure}[thbp!]
    \begin{center}
    \resizebox{85mm}{!}{\includegraphics{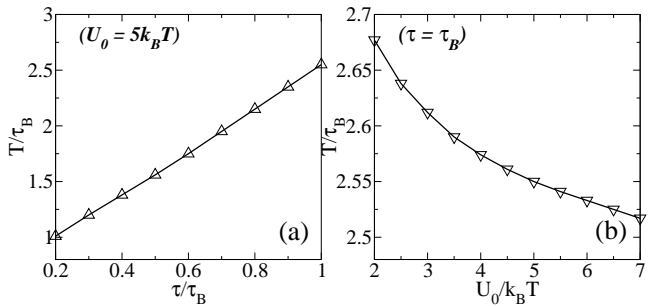}}
      %\resizebox{150mm}{!}{\includegraphics{circlefreq.eps}}
      \caption{Density oscillation cycle time $T$ (a) as a function of the delay time $\tau$ and (b) as a function of the washboard potential amplitude $U_0$ within the oscillatory regime [see Fig.~\ref{F2}]. The parameters are $F_0=7k_BT/\sigma$, $K_0=7k_BT/\sigma$ and $L=10\pi\sigma$. The rest of the parameters as in the legend.} %and $L=60\sigma$. \textit{We changed here to $U_0=5k_BT$, because we wanted a larger linearly unstable region.}}
      \label{Fig.circlefreq}
    \end{center}
\end{figure}
 In Fig.~\ref{Fig.circlefreq}(a) we plot the cycle time $T$ as a function of the delay time $\tau$. Clearly, by increasing the delay time $\tau$ the cycle time $T$ increases as well. To understand this behavior it is crucial to recall that the time-delayed feedback force incorporated here is of Pyragas type, i.e. a control target 
at time $t$ and its value at time $t-\tau$ is used for the feedback signal. In our case the control target is the centre of mass position $\langle z\rangle_t$, which is being shifted as a function of time due to the constant force $F^\text{bias}$. Thus, a larger delay time $\tau$ implies that the system travels longer distances within the time interval $\tau$. 
On the other hand, the specific form of the feedback force $F^\text{fb}$ is constructed such that it always counteracts $F^\text{bias}$ (see Eq.~(\ref{eq.fbforce}) and Refs. \cite{EPL40007,PhysRevE.79.041114}). Furthermore, the absolute value of $F^\text{fb}$ is small for large differences $\bar{f}(t)-\bar{f}(t-\tau)$. 
As a result, the crossover region where the feedback force $F^\text{fb}$ changes from being essentially inactive to compensating the constant tilting force $F^\text{bias}$ is accessed more often when the delay time is smaller. Thus, smaller delay times $\tau$ yield decreased cycle times $T$ up to the limit where $\tau$ is too small to induce time-periodic oscillations in the density any longer [see Fig.\ref{F2}(b)].

Fig.~\ref{Fig.circlefreq}(b) shows the dependence of $T$ on the washboard amplitude $U_0$. It is seen that the cycle time $T$ decreases slightly as a function of the washboard amplitude $U_0$. It is well known that the energy barrier plays a decisive role for the particle escape rate in a potential minimum for hopping processes that are thermally activated \cite{Kramers1940284}. In our case, the interplay between the washboard potential, the constant tilting force and the control force determines the rate at which particle are crossing to the next potential minimum. 
By increasing the washboard amplitude $U_0$, the energy barrier for a particle escape is increased leading to smaller displacements of the average particle position in a valley. As a result the cycle time of the oscillations is decreased.

\section{Microchannel geometry}\label{SecDYNMicro}
\begin{figure*}[tbp!]
    \begin{center}
    \resizebox{140mm}{!}{\includegraphics{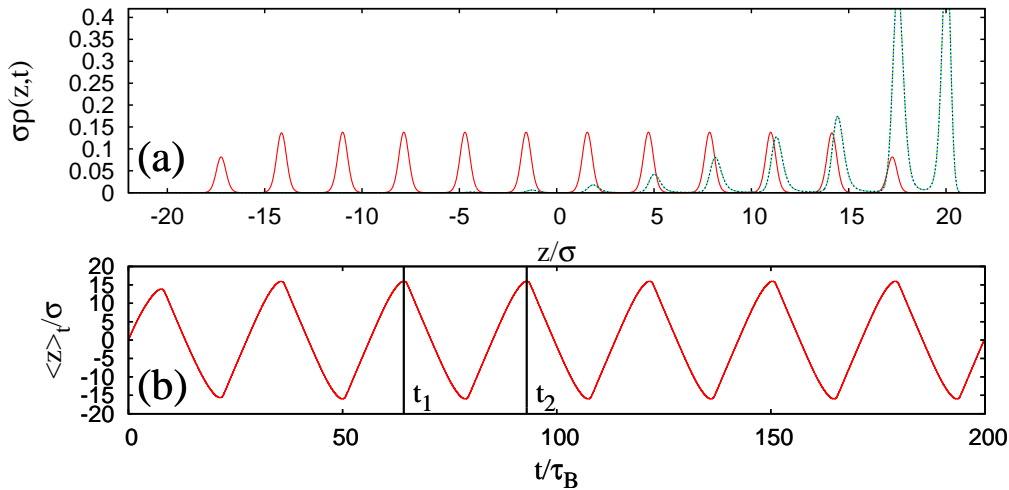}} %180
      %\resizebox{140mm}{!}{\includegraphics{U8K7F7micro.eps}} %180
      \caption{(Color online) Results for the microchannel, (a) density distribution $\rho(z,t)$ as a function of position for selected times $t_0/\tau_B=0$ (red curve), $t_1/\tau_B=64.3$ (green-dashed curve) and $t_2/\tau_B=93.0$ (blue-dashed curve) and (b) shows the first moment of $\rho(z,t)$ as a function of time. The parameters are $U_0=8k_BT$, $F_0=7k_BT/\sigma$, $K_0=7k_BT/\sigma$, $\tau=\tau_B$ and $z_\text{wall}=\pm 20\sigma$.}
      \label{fig.MicroDensityUnstable}
    \end{center}
\end{figure*}
So far we have considered infinite systems (i.e. systems with periodic boundaries). In this section we explore to which extent the emergence of an oscillation instability also depends on the channel geometry.
To this end, we now consider a system that consists of a microchannel of finite length $L$ with hard walls at the ends. We choose as initial distribution the equilibrium density distribution for a single particle subject to the washboard potential
plus a wall potential, $\beta U^\text{wall}=10(z/z_\text{wall})^{20}$, which confines the particle position to values $|z|\lesssim z_\text{wall}=20\sigma$. 
Such a smooth wall potential is typically used to model situations where the diameter of the particles forming the wall is much smaller than that of the
fluid particles \cite{grandner:244703}.
Thus, the initial distribution is given by
\begin{align}
\rho(z,t=0)=\rho_0\exp\left[-\beta U_0\cos^2\left(\dfrac{z}{\sigma}\right)-10\left(\dfrac{z}{z_\text{wall}}\right)^{20}\right].
\end{align}

Again, we focus on the ``balanced case" $\gamma=F_0-K_0=0$ with a fixed control amplitude value of $K_0=7k_\mathrm{B}T/\sigma$. We recall that in the case of the infinite system (periodic boundaries) a washboard amplitude value of $U_0=8k_BT$ is already sufficient
to suppress oscillatory states, yielding a stationary state for times larger than the transient response time. We choose this specific value for $U_0$ as a starting point for the microchannel study.

In Fig.~\ref{fig.MicroDensityUnstable}(a) we show snapshots of the density distribution at three different times $t_0=0\tau_B$, $t_1=64.3\tau_B$ and $t_2=93.0\tau_B$. We note that the initial density distribution $\rho(z,t=0)$ (shown as a red curve) is symmetric with respect to the position $z=0$. After a transient response time of roughly $20\tau_B$ we find stable time-periodic density oscillations where the distribution oscillates between two states characterized by a low and a high average particle position, respectively. Therefore, this state fulfills the periodicity condition $\rho(z,t+T)=\rho(z,t)$ with $T=2\pi/\omega$ being the cycle time of the oscillations.
The snapshots at times $t_1$ and $t_2=t_1+T$ (shown as a green-dashed and a blue-dashed curve in Fig.~\ref{fig.MicroDensityUnstable}(a), respectively), reveal that, indeed, both density profiles appear to be identical, at least to the naked eye. 
We support this conclusion by plotting in Fig.~\ref{fig.MicroDensityUnstable}(b) the average particle position $\langle z\rangle_t=\int_{0}^{L}z\rho(z,t)dz$ as a function of time. Here, the function $\langle z\rangle_t$ is averaged over the entire channel length $L$ in contrast to the circular ring geometry (periodic boundaries) where the obtained results are periodic with respect to each valley position. Similar as in Sec.~\ref{SecDYNNA}, the function $\langle z\rangle_t$ oscillates as a function of time with the same frequency $\omega$ as the frequency of the density oscillations, as can be seen from the time stamps $t_1$ and $t_2=t_1+T$ that we have included as vertical lines. However, we stress that the cycle times here are much longer than for the circular ring geometry. Furthermore, the periodic states have the spatial period equal to the largest spatial period used in the system, which is the system length itself as imposed by the wall potential. 
As a result, the time-periodic solution oscillates back and forth between $z=\pm z_\text{wall}$. 
We also note that the finding of density oscillations for $U_0=8k_BT$, $\gamma=0$ and $\tau=\tau_B$ is in contrast to what we found in Sec.~\ref{SecDYNNA} where no (periodic) instabilities occur for this specific parameter set.

By further increasing the washboard amplitude to the value of $U_0=9k_BT$, on the other hand, we find that the oscillatory behavior at early times is being damped resulting in a stationary, non-periodic density for times $t\gtrsim 20\tau_B$. 
We conclude that the results of the linear stabilty analysis (cf. Fig.~\ref{F2}) can approximately be used as a reference to find states that are linearly unstable for the microchannel system. We argue that this is because the channel is so large $(L=44\sigma)$ that the system is mostly determined by the bulk properties. The instability region for the finite channel seems to be qualitatively similar, but increased in size compared to the results for the circular ring geometry.\\ 
\begin{figure}[tbp!]
    \begin{center}
     \resizebox{85mm}{!}{\includegraphics{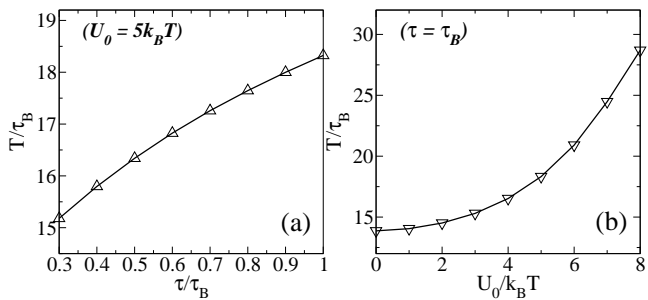}}
      %\resizebox{85mm}{!}{\includegraphics{microfreq.eps}}
      \caption{Density oscillation cycle time $T$ for the microchannel as (a) function of the delay time $\tau$ and (b) function of the washboard potential amplitude $U_0$. The parameters are $F_0=7k_BT/\sigma$, $K_0=7k_BT/\sigma$ and $z_\text{wall}=\pm 20\sigma$. The rest of the parameters as in the legend.} 
      \label{Fig.microTvstau}
    \end{center}
\end{figure}
In Fig.~\ref{Fig.microTvstau}(a) we show results for the cycle time $T$ as a function of the delay time $\tau$ for a fixed value of $K_0=7k_BT/\sigma$ and the choice $\gamma=F_0-K_0=0$. 
Holding the washboard potential amplitude fixed to the value of $U_0=5k_BT$, we do not find oscillatory density states below $\tau=0.3\tau_B$. By increasing $\tau$, we find a qualitatively similar behavior as in the periodic system (see Sec.~\ref{SecPerCycle}) for the cycle time $T$. Specifically, we find a monotonic increase on $T$ as a function of $\tau$. On the other hand, quantitatively comparing the results for the cycle time $T$ to the periodic system reveals that the values increase by approximately one order of magnitude. This is clearly a consequence of the substantial differences in the oscillations that we observe for the microchannel: the underlying nonlinear terms in the dynamical equations [see Eqs.~(\ref{eq.extpot}-\ref{eq:FPl1})] drive the density propagations over the whole system size, which is given by the channel length $L$. 
Thus, it is clear that the cycle time must be significantly longer than in the periodic system, where the oscillations occur around a valley position with displacements that are smaller than $\pi\sigma$ (the spatial period of the washboard potential). We note that for the $\tau$ values that we considered (up to $\tau=10\tau_B$), we do not find any upper boundary for the linear stability threshold. This behavior seems to be similar to the periodic system, where we showed that the oscillatory density state cannot be transformed into a stationary state by increasing the value of $\tau$ [cf. Fig.~\ref{F2}(b)].

In order to investigate the dependence of the cycle time $T$ on the washboard amplitude $U_0$, we hold the delay time fixed to the value of $\tau=\tau_B$ and increase (decrease) $U_0$ in steps of $\Delta U_0=1k_BT$ towards the linear stability threshold. We do not find oscillatory density states above $U_0^c\approx 8.63k_BT$. Furthermore, we observe a monotonic \textit{increase} of $T$ as a function of $U_0$, which contrasts with the periodic system where we found monotonic decrease. Again, this is a consequence of the substantially different oscillation mode that we observe for the microchannel.
%for $\gamma=0$ !!!
As explained above, any oscillatory perturbation applied to the system travels over the whole system length. Thus, an increased value of $U_0$ now means that the propagation of the perturbation is hindered, which results in an increased cycle time $T$ that corresponds to the travel time over a distance of approximately twice the system length $L$ (see Sec.\ref{SecPerCycle}).\\
\begin{figure}[tbp!]
    \begin{center}
      \resizebox{68mm}{!}{\includegraphics{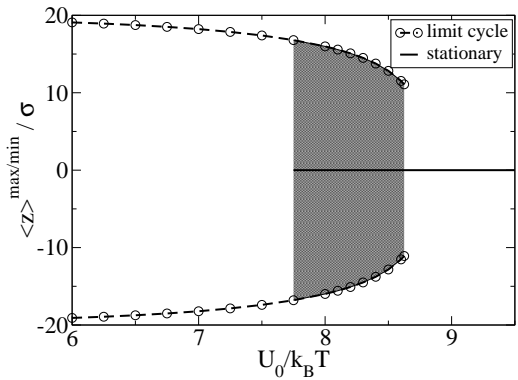}}
      %\resizebox{70mm}{!}{\includegraphics{bif_mc.eps}}
      \caption{Local extrema values of the function $\langle z\rangle_t$ as a function of the washboard amplitude $U_0$. The shaded area marks the region where the system exhibits hysteresis. The parameters are $F_0=7k_BT/\sigma$, $K_0=7k_BT/\sigma$, $\tau=\tau_B$, $z_\text{wall}=\pm 20\sigma$.}
      \label{Fig.circlemc}
    \end{center}
\end{figure}
For completeness, we show in Fig.~\ref{Fig.circlemc} the bifurcation diagram for the microchannel calculated in the same fashion as in Sec.~\ref{SecDYNNA}. We only find stationary states for washboard amplitudes $U_0>U_0^c$. Contrary to the infinite system, however, the average particle position $\langle z\rangle_t$ in this stationary state is principally a \textit{non-zero} constant as $t\rightarrow\infty$. For simplicity, all stationary states in Fig.~\ref{Fig.circlemc} have been shifted to zero. %Furthermore, the bifurcation point $U_0^c\approx 8.63k_BT$ is slightly shifted towards higher washboard amplitude values [cf. Fig.~\ref{Fig.circlebif}]. 
Upon increase of $U_0$ all density oscillations suddenly drop off at $U_0=U_0^c$ with a jump in the (oscillation) amplitude from the value $|\langle z\rangle^\text{max}|$ to zero; i.e., a \textit{subcritical} Hopf bifurcation occurs. 
As the parameter $U_0$ is reversed, a stationary solution can be found below the Hopf bifurcation point for values ranging to $U_0^m\approx 7.75k_BT$. Thus, the system system exhibits hysteresis within the parameter region $U_0^m<U_0<U_0^c$ (shown as the shaded area in Fig.~\ref{Fig.circlemc}).

\section{Influence of repulsive particle interactions}\label{SecPartInteract}
\begin{figure}[tbp!]
    \begin{center}
     \resizebox{70mm}{!}{\includegraphics{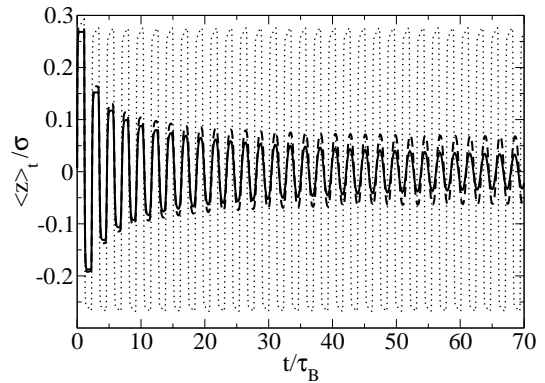}}
     % \resizebox{70mm}{!}{\includegraphics{interact.eps}}
      \caption{Results for the average particle position $\langle z\rangle_t$ as a function of time for different repulsion strengths: $\varepsilon_0=0k_BT$ (solid curve), $\varepsilon_0=12k_BT$ (dashed curve) and $\varepsilon_0=15k_BT$ (dotted curve). The parameters are $U_0=15k_BT$, $F_0=7k_BT/\sigma$, $K_0=7k_BT/\sigma$, $\tau=\tau_B$, $L=10\pi\sigma$ and $N=2$.}
      \label{Fig.interactions}
    \end{center}
\end{figure}
In many real colloidal systems, interactions between the particles cannot be neglected. Here, we briefly consider the case of purely repulsive interactions within the periodic system. To this end, 
we utilize the recently developed dynamical density functional theory (DDFT). The DDFT key equation is given by \cite{marconi:a413,marconi:8032,ArcherSpinDec}
\begin{align}
    \Gamma^{-1}\frac{\partial \rho(z,t)}{\partial t}=\nabla\cdot \left[\rho(z,t)\nabla \frac{\delta \mathcal{F}[\rho(z,t)]}{\delta \rho(z,t)}\right].\label{eq:1dDDFT}
\end{align}
The mobility coefficient in Eq.~(\ref{eq:1dDDFT}) is the same as in the Fokker-Planck approach [see Eq.~(\ref{eq:FPl1})], i.e. we can set its value to $\Gamma \tau_\mathrm{B}k_\mathrm{B}T/\sigma^2=1$.
The chemical potential $\mu(z,t)$ obtained from the Helmholtz free energy functional $\mathcal{F}$ has three contributions
\begin{align}
\mu(z,t)=\frac{\delta \mathcal{F}[\rho(z,t)]}{\delta \rho(z,t)}=\mu_\text{id}(z,t)+\mu_\text{int}(z,t)+\mu_\text{ext}(z,t)
\end{align}
The first contribution is the ideal gas term $\mu_{id}=k_BT\ln\Lambda \rho(z,t)$ ($\Lambda$ denotes the thermal de Broglie wavelength), the second contribution $\mu_\text{int}$ accounts for particle interactions and the third contribution is the external potential $\mu_\text{ext}=U_\text{ext}$, which in our case 
includes the contributions from the tilted washboard potential and the control force [see Eq.~(\ref{eq.extpot})].
In the following, the colloidal interactions are treated within a mean-field approach, that is, $\mu_\text{int}(z)=\int\negthickspace dz' \rho(z',t)U^\text{rep}(|z-z'|)$.
We employ 
the Gaussian Core Model (GCM) where the interaction potential is given by $U^\text{rep}(|z-z'|)=\varepsilon_0\exp\left[-(z-z')^2/\sigma^2\right]$. 
The GCM represents a typical coarse-grained potential which describes a wide class of \emph{soft} macroparticles with effective (gyration) radius $\sigma$ \cite{likos01, louis00}. We choose positive repulsion strengths $\varepsilon_0>0$, such that the interaction is purely repulsive.
Also, we focus in this section on the circular ring geometry (periodic boundaries) and consider the case of $N=2$.

We note that, even in the non-interacting case, the results from the linear stability analysis (see Sec.~\ref{SecLSA}) cannot be used as a basis for comparison. The reason is that the calculations in Sec.~\ref{SecLSA} were done with $N=1$. Rather, the washboard amplitude $U_0$ must be increased significantly to find a stable stationary state. For example, for the ``balanced case" $\gamma=F_0-K_0=0$ (and $\varepsilon_0=0$) we find oscillatory density states for a broad spectrum of values for $U_0$ ranging up to $U_0\approx 14k_BT$ (recall that $U_0=8k_BT$ is sufficient to suppress oscillatory states for $N=1$ and the rest of the parameters being the same).\\ 
In Fig.~\ref{Fig.interactions} we show results for the average particle position $\langle z\rangle_t$ for $U_0=15k_BT$ (and $N=2$). In the case of the non-interacting system $\varepsilon_0=0$ (shown as a solid curve in Fig.~\ref{Fig.interactions}) the function $\langle z\rangle_t$ is indeed being damped as a function of time, reflecting a stationary state for times larger than the transient response time, which is here of the order of $t_\text{res}\approx 2000\tau_B$.

Upon increase of the repulsion strength $\varepsilon_0$ the extrema of the function $\langle z\rangle_t$ increase (see the dashed curve and the dotted curve in Fig.~\ref{Fig.interactions}, respectively). 
This is consistent with our earlier finding \cite{EPL40007} that repulsive interactions support the particles in crossing the barrier, yielding an increase of the long-time diffusion coefficient.
Moreover, for $\varepsilon_0=15k_BT$ we find stable (time-periodic) density oscillations with cycle time $T=2.315\tau_B$. 
Thus, repulsive interparticle interactions can be successfully used to stabilize the oscillatory density state. We note, however, that for the present system increasing the average density (via the particle number $N$) has a considerably greater impact on the linear stability of the system.

\section{Concluding remarks}
In the present work, we have investigated the dynamics of colloidal particles subject to a one-dimensional, tilted washboard potential under time-delayed feedback control.
The major goal was to identify conditions under which the control can induce oscillatory states. The latter are absent in the uncontrolled system. Our investigations
are based on the (nonlinear) Fokker-Planck equation, combined with a linear stability analysis.

We have investigated infinite systems (i.e., systems with periodic boundaries) and microchannels of finite length L (bounded by repulsive walls). For the first case, we have obtained, based on linear stability analysis, a full state diagram. This diagram predicts that oscillations (of the density field) do indeed occur for finite values of the delay time and control strengths comparable to the strengths of the conservative forces. 
Investigating the time dependence of the same system via numerical solution of the Fokker-Planck equation, we found 
full consistency with the results of the linear stability analysis for all model parameters considered. In addition, the numerical solution provides results for the (likewise oscillating) moments of the density distribution. In particular, from the
oscillations of the first moment (i.e., the current) we could identify the cycle time. The latter was found to monotonically decrease with the delay time.

In the finite system (which we have investigated via the full Fokker-Planck equation alone), we also found oscillations. However, the spatial extension of these oscillations corresponds to the wall-to-wall separation, rather than to the width of one valley, as in the infinite-system case. Despite these differences, the parameter region where the finite system exhibits feedback-induced oscillations is rather similar to that found in the infinite case. We attribute this to the fact that the finite system under consideration was still so large
($L\gg \lambda_\text{wb}=\pi\sigma$), that boundary effects are not dominant.  
Finally, we have briefly considered the case that the colloidal particles interact. 
This was done on the basis of the recently developed DDFT \cite{marconi:a413,marconi:8032,ArcherSpinDec},
a generalized continuity equation where the particle interactions enter via a free energy functional. 
Using a purely repulsive (GCM) pair potential, we have shown that such interactions can {\it stabilize} 
oscillatory states. However, to see this effect the strength of repulsion must be of the order of the washboard amplitude (and both must be significantly larger than $k_{\mathrm{B}}T$).
Taken together, we have shown that colloidal particles in modulated potentials under time-delayed feedback control can display highly non-trivial dynamics, particularly oscillations. 
One way to understand these differences to the uncontrolled case is that our feedback control term, which relies on the average particle position and thus involves {\it all} particles, introduces effectively time-dependent interactions between the particles.

We note that feedback-induced spatio-temporal effects have been recently also been found in other extended systems such as optical resonators, where the delayed feedback generates
spontaneous motion of cavity solutions \cite{PhysRevLett.103.103904}, and, more generally, systems describable by the Swift-Hohenberg equation \cite{Gurevich/arXiv}. A particularly interesting feature of the present colloidal system is that it is, in principle, accessible by experiments \cite{PhysRevE.77.041107,PhysRevLett.93.040603,PhysRevLett.106.168104}. Indeed, colloidal particles are typically so large that their position (and, consequently, the first moment of the density distribution) can be easily monitored by real-space methods such as video microscopy. We therefore hope that our results will stimulate
future experiments. Moreover, from the theoretical side one could extend the present analysis to mixtures consisting of several species with multiple time delay constants. This
 could be a promising route for the development of a novel
particle sorting effect in narrow channels.
%Moreover, so far we have considered only a small portion of the parameter space. Indeed, we have focussed on the ``balanced case" where the amplitudes of the constant tilting force and the feedback force are the same, that is  $\gamma=F_0-K_0=0$.
%We note that for $\gamma\neq 0$ the spatial period of the density oscillations (corresponding to the distance the system travels over a period) can be much smaller. {\bf \color{red} Remark: I don't understand what you are trying to say here. The spatial period is always fixed and equal to either the period of the washboard potential or the length of the channel.}
%This phenomenon might be of interest for technical applications such as particle separation techniques in narrow channels \cite{PhysRevLett.108.020604}. Specifically, extending the present system to mixtures consisting of several species with multiple time delay constants could be a promising route to the development of a novel particle sorting effect. Work in this direction is under way.

\acknowledgements

We gratefully acknowledge financial support via the Collaborative Research Center (SFB) 910 ``Control of self-organizing nonlinear systems: Theoretical methods and concepts of application" and the Research Training Group (GRK) 1558
``Nonequilibrium Collective Dynamics in Condensed Matter and Biological Systems".

\begin{appendix}
\section{Numerical continuation of solutions of the linearized Fokker-Planck equation}
\label{append}
It is known that in order to continue any given solution of the below Sec.~\ref{SecLSA} mentioned boundary value problem (BVP) with 7 equations,
$7$ boundary conditions and $3$ integral conditions, one requires $(7+3-7+1 = 4)$ continuation parameters. We always include the pair $(\omega,\lambda)$ into the set of continuation parameters.
Consequently, we are left only with two additional continuation parameters. These can be
chosen in the arbitrary fashion out of the set $(U_0,K_0, F_0,\tau, N)$. For example, the solution
of the BVP can be continued in the plane of $(U_0,K_0)$, with all other parameters fixed.
The result of such continuation is a line in the space $(U_0,K_0)$. Generally, any continuation
yields a co-dimension one manifold in the space of system parameters, which represents the
stability threshold. The onset frequency $\omega$ is then an output parameter, which is given
by a certain function of the position on the threshold. The accuracy of all continuation
runs is controlled by checking that the absolute value of the eigenvalue $\lambda$ is of the order of $10^{-10}\dots 10^{-12}$.\\
Generally, it is not easy to find a particular combination of the system parameters directly
on the stability threshold, and additionally to correctly guess the corresponding value of $\omega$.
In order to find a starting point for our continuations, we employ the following strategy.
First, we switch off the integral conditions Eqs.\,(\ref{int_cond}) and continue an analytically known
solution at vanishing driving force ($F_0=0$), that is,
\begin{align}
\label{start_stat}
\rho_s(z)\sim e^{-U(z)},\,\,\,C(z)=0,\,\,\,\,S(z)=0,
\end{align}
with an arbitrary guess of $\tau$ and $\omega$ in parameter $\omega$, until the point $\langle S\rangle=0$ is found. Second,
we continue this solution in parameter $K_0$, until the point on the stability threshold is found,
i.e. $\langle S\rangle=0$ and $\langle C\rangle=1$. Note that the condition $\langle S\rangle=0$ remains uneffected, when continuing in the parameter $K_0$. Finally, we switch the integral conditions Eqs.\,(\ref{int_cond}) back on and proceed
with the continuation in any of the parameters $(U_0,K_0,F_0,\tau,N)$, as described above.
\end{appendix}

%\addcontentsline{toc}{chapter}{\\ \\ \large Literaturverzeichnis}
%\bibliography{paper}
%

\end{document}